\documentclass[twocolumn,aps,prl,showpacs,preprintnumbers,amsmath,amssymb, superscriptaddress,longbibliography]{revtex4-1}
\pdfoutput=1
\usepackage{bm}
\usepackage{graphicx}
\usepackage{color,xspace}
\begin{document}

\title{Observation of high-$T_c$ superconductivity in rectangular FeSe/STO(110) monolayer}

\author{P. Zhang}\thanks{These authors contributed equally to this work.}
\affiliation{Beijing National Laboratory for Condensed Matter Physics, and Institute of Physics, Chinese Academy of Sciences, Beijing 100190, China}
\author{X.-L. Peng}\thanks{These authors contributed equally to this work.}
\affiliation{Beijing National Laboratory for Condensed Matter Physics, and Institute of Physics, Chinese Academy of Sciences, Beijing 100190, China}
\author{T. Qian}\thanks{These authors contributed equally to this work.}
\affiliation{Beijing National Laboratory for Condensed Matter Physics, and Institute of Physics, Chinese Academy of Sciences, Beijing 100190, China}
\author{P. Richard}
\affiliation{Beijing National Laboratory for Condensed Matter Physics, and Institute of Physics, Chinese Academy of Sciences, Beijing 100190, China}
\affiliation{Collaborative Innovation Center of Quantum Matter, Beijing, China}
\author{X. Shi}
\affiliation{Beijing National Laboratory for Condensed Matter Physics, and Institute of Physics, Chinese Academy of Sciences, Beijing 100190, China}
\author{J.-Z. Ma}
\affiliation{Beijing National Laboratory for Condensed Matter Physics, and Institute of Physics, Chinese Academy of Sciences, Beijing 100190, China}
\author{B.-B. Fu}
\affiliation{Beijing National Laboratory for Condensed Matter Physics, and Institute of Physics, Chinese Academy of Sciences, Beijing 100190, China}
\author{Y. L. Guo}
\affiliation{Beijing National Laboratory for Condensed Matter Physics, and Institute of Physics, Chinese Academy of Sciences, Beijing 100190, China}
\author{Z. Q. Han}
\affiliation{Beijing National Laboratory for Condensed Matter Physics, and Institute of Physics, Chinese Academy of Sciences, Beijing 100190, China}
\affiliation{Department of Physics, Renmin University, Beijing, 100872, China}
\author{S. C. Wang}
\affiliation{Department of Physics, Renmin University, Beijing, 100872, China}
\author{L. L. Wang}
\affiliation{State Key Laboratory of Low-Dimensional Quantum Physics, Department of Physics, Tsinghua University, Beijing 100084, China}
\affiliation{Collaborative Innovation Center of Quantum Matter, Beijing, China}
\author{Q.-K. Xue}
\affiliation{State Key Laboratory of Low-Dimensional Quantum Physics, Department of Physics, Tsinghua University, Beijing 100084, China}
\affiliation{Collaborative Innovation Center of Quantum Matter, Beijing, China}
\author{J. P. Hu}
\affiliation{Beijing National Laboratory for Condensed Matter Physics, and Institute of Physics, Chinese Academy of Sciences, Beijing 100190, China}
\affiliation{Collaborative Innovation Center of Quantum Matter, Beijing, China}
\affiliation{Department of Physics, Purdue University, West Lafayette, Indiana 47907, USA}
\author{Y.-J. Sun}\email{yjsun@iphy.ac.cn}
\affiliation{Beijing National Laboratory for Condensed Matter Physics, and Institute of Physics, Chinese Academy of Sciences, Beijing 100190, China}
\author{H. Ding}\email{dingh@iphy.ac.cn}
\affiliation{Beijing National Laboratory for Condensed Matter Physics, and Institute of Physics, Chinese Academy of Sciences, Beijing 100190, China}
\affiliation{Collaborative Innovation Center of Quantum Matter, Beijing, China}

\date{\today}

 %74.70.Xa, 74.25.Jb, 79.60.-i}

%74.70.Xa 	Pnictides and chalcogenides
%74.25.Jb 	Electronic structure (photoemission, etc.)
%79.60.-i 	Photoemission and photoelectron spectra
%73.20.At 	Surface states, band structure, electron density of states
%74.78.-w 	Superconducting films and low-dimensional structures

%\end{minipage}
\maketitle
%\narrowtext

\textbf{It is well known that superconductivity in Fe-based materials is favoured under tetragonal symmetry, whereas competing orders such as spin-density-wave (SDW) and nematic orders emerge or are reinforced upon breaking the fourfold (C$_4$) symmetry \cite{Fernandes_NatPhys2014}. Accordingly, suppression of orthorhombicity below the superconducting transition temperature ($T_c$) is found in underdoped compounds \cite{Nandi_PRL104}. Epitaxial film growth on selected substrates allows the design of crystal specific lattice distortions. Here we show that despite the breakdown of the C$_4$ symmetry induced by a $\sim$ 5\% difference in the lattice parameters, monolayers of FeSe grown by molecular beam epitaxy (MBE) on the (110) surface of SrTiO$_3$ (STO) substrates [FeSe/STO(110)] exhibit a large nearly-isotropic superconducting (SC) gap of 16 meV closing around 60 K. Our results on this new interfacial material, similar to those obtained previously on FeSe/STO(001) \cite{LiuNC2012,ZhouNM2013,TanNM2013,ShenNATURE2014}, contradict the common belief that the C$_4$ symmetry is essential for reaching high $T_c$'s in Fe-based superconductors.}

Since it has the simplest structure of all Fe-based superconductors, FeSe is a perfect candidate to study the effect of crystal distortions on superconductivity. The $T_c$ of bulk FeSe jumps from 8 K to 37 K upon suppressing an orthorhombic distortion by applying external pressure \cite{Margadonna_PRB80,NishigoriJPSJ2014}. Similarly, the recently discovered \cite{XueCPL2012} FeSe monolayers grown by molecular beam epitaxy (MBE) on SrTiO$_3$ (001) substrates [STO(001)], which have the highest $T_c$ (65 K) amongst all Fe-based materials, are characterized by a C$_4$ axis. Here we show a new approach to create a distortion in FeSe or other Fe-based superconductors.

\begin{figure*}
\begin{center}
\includegraphics[width=\textwidth]{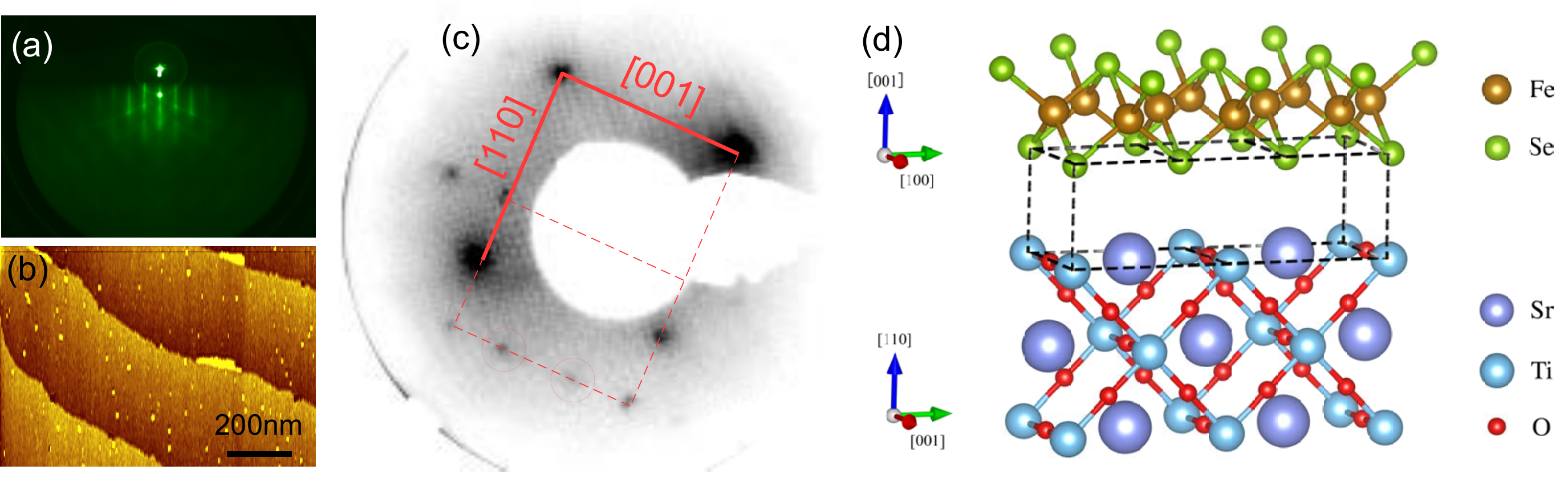}
\end{center}
\caption{\label{structure}\textbf{Characterization of the FeSe monolayer films.} \textbf{a}-\textbf{b} RHEED pattern and STM image of a FeSe/STO(110) film, respectively. \textbf{c} LEED pattern of FeSe/STO(110) at 170 K. The BZ of the FeSe film is indicated by solid lines, while dashed lines indicate the BZ of the STO substrate projected on the (110) surface. The red circles correspond to well-known $3\times 1$ surface reconstruction on STO(110) surface \cite{CastellPRB2008}. \textbf{d} Crystal structure of a FeSe monolayer on top of a STO(110) substrate. The dashed lines indicate the possible match between the FeSe film and the STO(110) substrate.}
\end{figure*}

In contrast to STO(001) substrates, which expose a tetragonal surface of lattice parameter $a=3.905$ \AA\, for growth, STO(110) exposes a surface with characteristic lengths $a$ and $\sqrt{2}a$. The huge lattice mismatch between the STO(110) surface and FeSe (3.765 \AA\, \cite{Hsu_PNAS105}) does not allow a one-unit-cell to one-unit-cell growth as in the case of FeSe/STO(001). The LEED pattern of our samples [Fig. \ref{structure}c] indicates clearly an orthorhombic pattern that is consistent with 3 unit cells of FeSe growing on the top of 2 STO(110) unit cells, as illustrated in Fig. \ref{structure}d. Consequently, this means that the FeSe monolayer is characterized by lattice parameters $a$ and $\frac{2}{3}\sqrt{2}a=0.94a$. As compared to tetragonal FeSe/STO(001), this orthorhombic distortion in FeSe/STO(110) leads to a 6\% compression along one axis corresponding to the in-plane projection of the Fe-Se bonding ([110] direction), in agreement with our analysis of the LEED pattern, which indicates an anisotropy of $\sim 5\%$ .

\begin{figure*}
\begin{center}
\includegraphics[width=\textwidth]{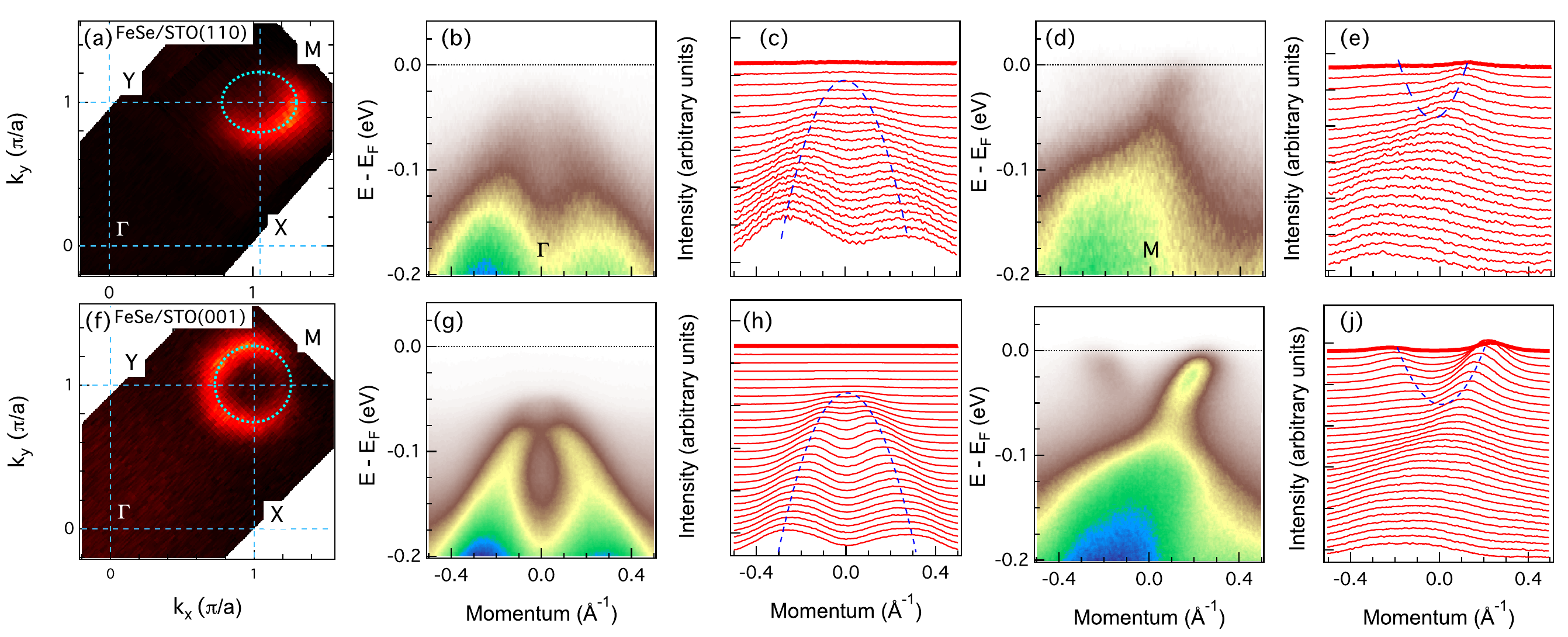}
\end{center}
\caption{\label{band}\textbf{Comparison between the electronic structure of monolayer films of FeSe/STO(110) and FeSe/STO(001).} The top and bottom rows correspond to results obtained on monolayer films of FeSe/STO(110) and FeSe/STO(001), respectively. The data is recorded at 35K. \textbf{a} and  \textbf{f} Fermi surface. The blue curves indicate the FS. \textbf{b} and  \textbf{g} ARPES intensity cut along a cut passing through the $\Gamma$ point. The dotted blue line is a parabolic fit to the band dispersion. \textbf{c} and  \textbf{h} Same as \textbf{b} and  \textbf{g} but for a cut passing through M.}
\end{figure*}

The large lattice distortion induced by the STO(110) substrate should have a significant impact on the electronic structure of the FeSe films. The compression of the unit cell enlarges the Brillouin zone (BZ) along M-Y and consequently elongates the M-centred electron pockets along that same direction. To check if this is really the case, we performed angle-resolved photoemission spectroscopy (ARPES) measurements on both FeSe/STO(110) and FeSe/STO(001). The comparison of the electronic band structures is shown in Fig. \ref{band}. As previously reported \cite{LiuNC2012,ZhouNM2013,TanNM2013,ShenNATURE2014}, the Fermi surface (FS) of FeSe/STO(001) consists in one doubly degenerate circular electron pocket centred at M. Consequently, the samples are highly-electron-doped compared to bulk FeSe single-crystals, for which both hole and electron pockets are measured by ARPES \cite{Maletz_PRB89,Shimojima_PRB90,Watson_PRB91,P_Zhang_PRB91}. In contrast to FeSe/STO(001), the FS of FeSe/STO(110) is not circular but rather elliptical, with the long axis of the ellipse parallel to M-Y, rather than along $\Gamma$-M like in the other Fe-based superconductors. Interestingly, the distortion of the ellipse is larger than the orthorhombic distortion of the lattice. Indeed, we find $\sim$ 0.8 for the ratio of two axes of the ellipse.

%%%%%%%%%%%%%%%%%%%%%%%%%%%%%%%%%%%%%%%%%%%%%%%%%
%  Figure 1
%%%%%%%%%%%%%%%%%%%%%%%%%%%%%%%%%%%%%%%%%%%%%%%%%

The FS of our FeSe/STO(110) and FeSe/STO(001) samples do not differ only by the shape, but also by the size, the latter one being larger. According to the Luttinger theorem, this indicates a smaller electron doping level in FeSe/STO(110). Assuming that each FS is degenerate, we estimate that the doping level of FeSe/STO(110) is 4.3\%/Fe, whereas it is about 5.5\% for our FeSe/STO(001) samples. This difference in the doping levels is also evident from the band dispersions. Indeed, the hole band at $\Gamma$ and the electron band at M are shifted towards higher binding energies ($\sim$ 30 meV at $\Gamma$ and $\sim$ 10 meV at M) in FeSe/STO(001) as compared to FeSe/STO(110). One possible explanation for the difference of doping is the smaller number of oxygen atoms exposed on the STO(110) surface as compared to the STO(001) surface, which may affect the electron transfer caused by oxygen vacancies on the surface of the substrate.

%%%%%%%%%%%%%%%%%%%%%%%%%%%%%%%%%%%%%%%%%%%%%%%%%
%  Figure 2
%%%%%%%%%%%%%%%%%%%%%%%%%%%%%%%%%%%%%%%%%%%%%%%%%

%%%%%%%%%%%%%%%%%%%%%%%%%%%%%%%%%%%%%%%%%%%%%%%%%
%  Figure 3
\begin{figure*}
\begin{center}
\includegraphics[width=\textwidth]{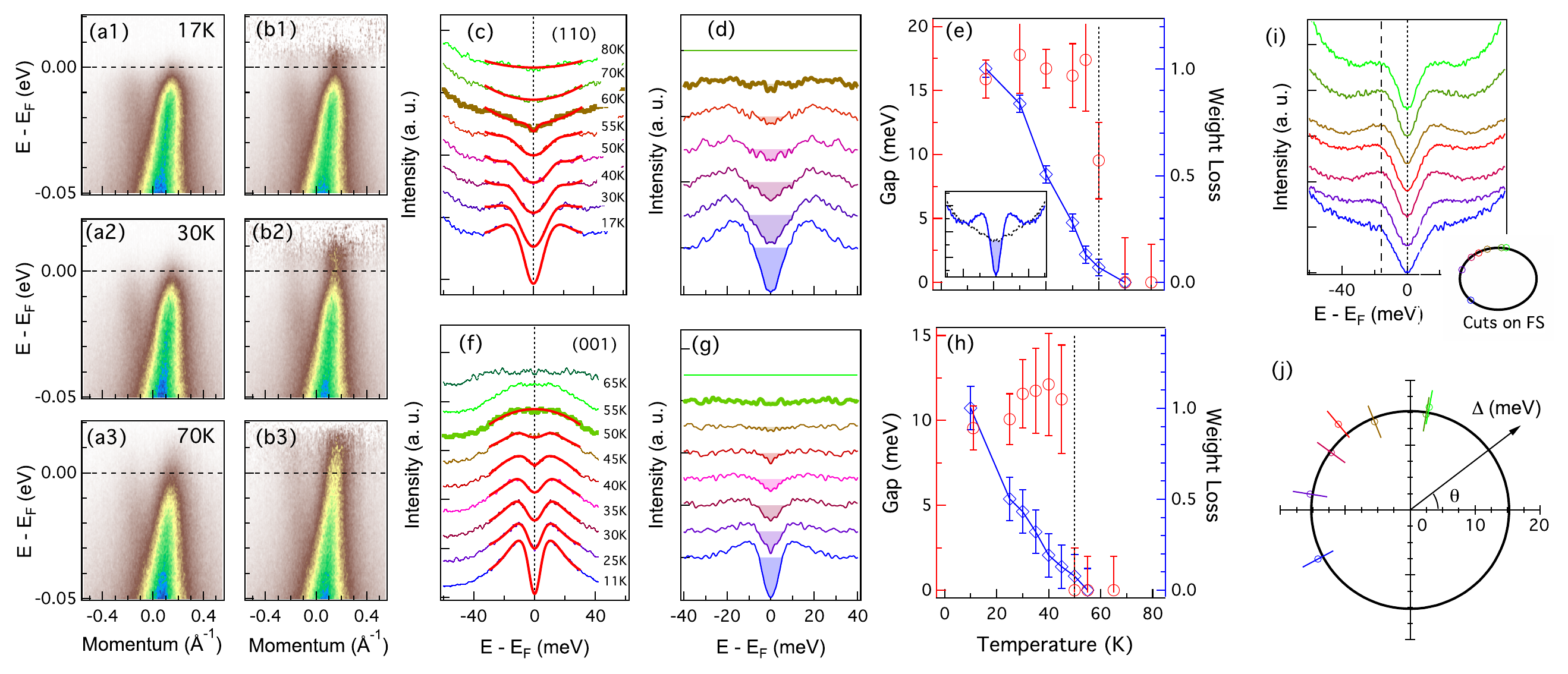}
\end{center}
 \caption{\label{gapt}\textbf{Superconducting gap.} \textbf{a1 - a3} Band structure of FeSe/STO(110) at M recorded at different temperatures. \textbf{b1 - b3} Same as \textbf{a1 - a3}, but divided by the Fermi-Dirac function. \textbf{c} and \textbf{f} Temperature evolution of the symmetrized EDCs at $k_F$ of FeSe/STO(110) and FeSe/STO(001), respectively. The thick red curves are fitting results using a model described in the text. \textbf{d} and \textbf{g} Corresponding symmetrized EDCs after subtraction of a high temperature background. The shadowed areas correspond to the difference between the background and the data, as illustrated in the inset of panel \textbf{e}. \textbf{e} and \textbf{h} Gap size and spectral weight loss as a function of temperature in FeSe/STO(110) and FeSe/STO(001), respectively. The gap is obtained from the fits shown in \textbf{c} and \textbf{f}. \textbf{i} Momentum distribution of the EDCs recorded at 17 K on the FS of FeSe/STO(110) and \textbf{j} corresponding polar representation of the SC gap amplitude. The inset of \textbf{i} shows the momentum locations of the EDCs.}
\end{figure*}
%%%%%%%%%%%%%%%%%%%%%%%%%%%%%%%%%%%%%%%%%%%%%%%%%

In Figs. \ref{gapt}a1-a3, we show the temperature evolution of the band structure of our FeSe/STO(110) monolayer samples. Changes mainly occur near the Fermi level ($E_F$). At 17 K, we see a gap feature that becomes more obvious after dividing the spectrum by the Fermi-Dirac function [Fig. \ref{gapt}b1]. No such gap is observed at 70 K [Fig. \ref{gapt}b3]. To extract the gap size, we followed the common practice consisting in symmetrizing the energy distribution curves (EDCs) at the Fermi wave vector ($k_F$), which removes the Fermi-Dirac cutoff, as shown in Fig. \ref{gapt}c. We normalized all the symmetrized EDCs from -0.1 eV to 0.1 eV to satisfy spectral weight conservation. We further subtracted the data at 70 K, used as a background. The results are shown in Fig. \ref{gapt}d. Since the coherent peak becomes weaker and broader with temperature increasing, the gap size $\Delta$ is difficult to extract precisely. We thus used a phenomenological model to extract it \cite{Norman_PRB57}, in which we consider the self-energy $\Sigma(\mathbf{k},\omega)=-i\Gamma_1+\Delta^2/[(\omega+i0^{+}+\varepsilon(\mathbf{k})]$, where $\omega$ is the energy, $\varepsilon(\mathbf{k})$ is the electronic band dispersion and $\Gamma_1$ is a constant single-particle scattering rate. The spectral function $A(\mathbf{k},\omega)$ is defined by $\pi A(\mathbf{k},\omega)= \Sigma^{\prime\prime}(\mathbf{k},\omega)/[(\omega-\varepsilon(\mathbf{k})-\Sigma^{\prime}(\mathbf{k},\omega))^2+\Sigma^{\prime\prime}(\mathbf{k},\omega)^2]$, where $\Sigma^{\prime}$ and $\Sigma^{\prime\prime}$ are the real and imaginary parts of the self-energy, respectively, and $\varepsilon(\mathbf{k}_F)=0$. After convoluting the spectral function by the instrumental resolution function, we can fit the data at $\mathbf{k}=\mathbf{k}_F$ and extract the value of $\Delta$, which is contained into $\Sigma^{\prime}$. The results are displayed in Figs. \ref{gapt}e and h, for FeSe/STO(110) and FeSe/STO(001), respectively. A nearly-temperature-independent gap size of about 16 meV is found in FeSe/STO(110) below $T_c$ (Fig. \ref{gapt}e). For FeSe/STO(001) we see a small temperature dependence of the gap, which is about 10 meV at the lowest temperature. %, where the fit should be more reliable.

Because the gap looks more filling than closing with temperature, we also used data from Fig. \ref{gapt}d to extract the spectral weight loss, defined as the area between the high-temperature background (70 K) and the symmetrized EDCs, which is also equal, below $T_c$, to the weight of the coherent peak. Similar procedure was successfully used in the past \cite{Kondo_Nature457} to investigate the SC gap and the pseudogap in (Bi,P)$_2$(Sr,La)$_2$CuO$_{6+\delta}$. The method is illustrated in the inset of Fig. \ref{gapt}e. For both FeSe/STO(110) and FeSe/STO(001), the spectral weight loss decreases with temperature increasing, and it vanishes around 60 K and 50 K, respectively, which are consistent to the temperatures where the gap size closes in our fits. We can thus use these values as reliable definitions of $T_c$. Keeping in mind that the gap size changes with the annealing conditions in FeSe/STO(001) \cite{ZhouNM2013}, our results on FeSe/STO(110) are very similar, both qualitatively and quantitatively, to results obtained on FeSe/STO(001), also shown in Fig. \ref{gapt}. The gaps in both materials thus have the same origin and hereafter, following previous reports on FeSe/STO(001) \cite{LiuNC2012,ZhouNM2013,TanNM2013,ShenNATURE2014}, we will refer to them as SC gaps.

%The latest results are quite surprising considering that a C$_2$ distortion is believed to be detrimental to superconductivity.
To check how the gap size evolves along the anisotropic FS of FeSe/STO(110), we measured the gap amplitude at different momentum locations along the FS, as shown in the inset of Fig. \ref{gapt}i. Surprisingly, we observe a rather isotropic gap, as illustrated by the polar distribution of the gap in Fig. \ref{gapt}i. In other words, not only superconductivity with a large SC gap survives to the C$_2$ lattice distortion, there is no obvious imprint of the lattice distortion on the gap distribution. As a corollary, the C$_4$ symmetry is not essential for reaching high-$T_c$ values in Fe-based superconductors.

\begin{figure}
\begin{center}
\includegraphics[width=\columnwidth]{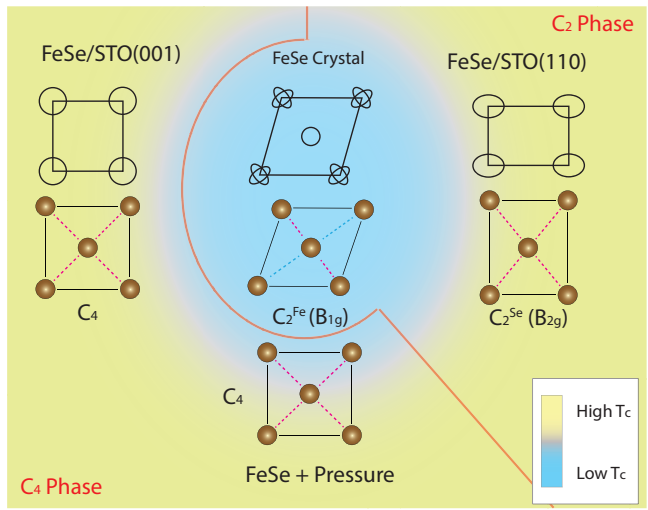}
\end{center}
 \caption{\label{Fig_distortion}\textbf{Various crystal structures of superconducting FeSe.} Superconductivity in FeSe is found in different forms, illustrated here (Fe atoms only) along with the corresponding FSs. The relative $T_c$ of the materials is given by the colour of the background. The dashed colour lines indicate which Fe-Fe bondings are equivalent. FeSe with C$_4$ symmetry is found in monolayer films of FeSe/STO(001) \cite{XueCPL2012}, as well as in single-crystals under hydrostatic pressure \cite{Margadonna_PRB80,NishigoriJPSJ2014}. While the FS of the monolayer films of FeSe/STO(001) consists in a doubly-degenerate circular electron pocket at the M point \cite{LiuNC2012,ZhouNM2013,TanNM2013,ShenNATURE2014}, the FS of the single-crystal under pressure is unknown. The single-crystals of FeSe suffer a small C$_2^\textrm{Fe}$ distortion (B$_{1g}$) leading to inequivalent Fe-Fe bondings. The FS of the single-crystals if formed by a $\Gamma$-centred hole pocket and two perpendicularly aligned M-centred electron pockets with ellipsoidal shapes. Finally, monolayer films of FeSe/STO(110) exhibit a C$_2^\textrm{Se}$ distorsion (B$_{2g}$) that keeps all Fe-Fe bondings equivalent. Their FS consists in a doubly-degenerate ellipse elongated along M-Y.}
\end{figure}

The different SC crystal structures of FeSe, with the corresponding schematic FSs, are summarized in Fig. \ref{Fig_distortion}. The C$_2$ distortion observed in FeSe/STO(110), which we hereafter refer to as C$^{\textrm{Se}}_2$ distortion (B$_{2g}$ irreducible representation for the square lattice with Fe atoms only), is different from the C$^{\textrm{Fe}}_2$ distortion (B$_{1g}$) observed in bulk FeSe \cite{NishigoriJPSJ2014} and in the parent compounds of many ferropnictide families \cite{Rotter2008,de_la_Cruz_Nature2008,Z_Li_PRB80,SL_Li_PRB80}. In the latter case, the distortion induces inequivalent Fe-Fe bondings in perpendicular directions, which stabilizes the long-range stripe spin-density-wave order in BaFe$_2$As$_2$ \cite{Huang_PRL101} and the long-range orbital ordering in bulk FeSe \cite{NishigoriJPSJ2014}. Since these long-range orderings are associated to order parameters competing with superconductivity \cite{Fernandes_NatPhys2014}, the C$^{\textrm{Fe}}_2$ distortion is clearly an obstacle for achieving high-$T_c$ superconductivity. However, no competing electronic order  associated with the C$^{\textrm{Se}}_2$ distortion has been observed. Therefore, our results strongly indicate that the lattice degree of freedom itself does not directly affect superconductivity. The SC pairing mechanism is tied to electronic degrees of freedom, like spin \cite{MazinPRL2008,HuPRL2008} and orbital \cite{KontaniPRL2008}. We note that it has been proposed that when different SC order parameters have comparable free energies, the breakdown of the C$_4$ symmetry may increase $T_c$ by mixing these SC order parameters, thus lifting the pairing frustration \cite{Fernandes_PRL111}.

%%%%%%%%%%%%%%%%%%%%%%%%%%%%%%%%%%%%%%%%%%%%%%%%%
%  Figure 4

%%%%%%%%%%%%%%%%%%%%%%%%%%%%%%%%%%%%%%%%%%%%%%%%%

In a previous ARPES report \cite{ShenNATURE2014}, the observation of band replica has been attributed to a strong coupling with a surface phonon of STO(001), which could play an important role in boosting the $T_c$ of FeSe/STO(110). Although we did not observe such replica in our FeSe/STO(110) samples, we can neither exclude nor prove that a similar mechanism occurs, but explaining why FeSe/STO(110) and FeSe/STO(001) have similar $T_c$'s and gap sizes would require the ``boosting phonons" to involve the same vibrations of atoms along the surface of the substrate. Nevertheless, strong empirical conclusions emerge from our comparative work on FeSe/STO(110) and FeSe/STO(001) monolayer films: (1) A large electron doping is necessary for achieving a high-$T_c$ in ferrochalcogenide materials, possibly by pushing the $\Gamma$-centred hole pockets below $E_F$. Indeed, superconductivity has been achieved in 3-monolayer-thick samples of FeSe only after electron-doping their surface with K \cite{TakahashiNM2015}. (2) The C$_4$ symmetry is not necessary for achieving high-$T_c$ and superconductivity is not harmed by a C$^{\textrm{Se}}_2$ distortion that breaks rotational symmetry but leaves all Fe-Fe bondings equivalent. Our ARPES observation of high-$T_c$ superconductivity with an isotropic gap in this new type of interfacial material clearly challenges some of the current beliefs and sheds new lights into the mechanism of Fe-based superconductors.

\textbf{Methods} Monolayer films of FeSe were grown and measured in the MBE-ARPES combined system at the Institute of Physics, Chinese Academy of Sciences. Atomically flat 0.7 wt\% Nb-doped STO(110) substrates were obtained after annealing for 6 hours at 600$^{\circ}$C and then 1.5 hour at 950$^{\circ}$C. The substrates were kept at 310$^{\circ}$C during the film growth. Fe (99.98\%) and Se (99.999\%) were co-evaporated from Knudsen cells with a flux ratio of ~1:10, which were measured by a quart crystal balance, with the growth rate of 0.35 UC/min. After growth, the FeSe monolayer films were annealed at 365$^{\circ}$C for 8h to improve crystallinity, and subsequently transferred \emph{in situ} into the ARPES chamber. Refection high-energy electron diffraction (RHEED) pictures of the FeSe monolayer films (Fig. \ref{structure}a) and scanning tunnelling microscopy (STM) topographic images (Fig. \ref{structure}b), suggest high-quality films. The base pressure was $3\times 10^{-10}$ torr in the MBE chamber and better than $2\times 10^{-11}$ torr during the ARPES measurements. ARPES measurements were performed in the same combined system using a R4000 analyser with a helium discharge lamp, and at the ``Dreamline" beamline of the Shanghai Synchrotron Radiation Facility (SSRF) using a VG-Scienta D80 electron analyser. The energy resolution was set to $\sim$ 5 meV for gap measurements and $\sim$ 15 meV for the band structure and FS mapping. The angular resolution was set to 0.2$^\circ$. Monolayer films of FeSe/STO(110) have also been grown at Tsinghua University using the same method \cite{Tsinghua_STM}. The high-resolution STM spectra obtained on these samples are consistent with the ARPES presented in the current work.

%%%%%%%%%%%%%%%%%%%%%%%%%%%%%%%%%%%%%%%%%%%%%%%%%%%%

We acknowledge R. M. Fernandes, S. L. Li, S. Uchida, F. Wang, X. X. Wu, H. Yao and F. C. Zhang for useful discussions, as well as L.-Y. Kong, W.-L. Zhang and Y. Zou for technical assistance. This work was supported by grants from MOST (2015CB921000, 2010CB923000, 2011CBA001000, 2011CBA00102, 2012CB821403 and 2013CB921700) and NSFC (11574371, 11004232, 11034011/A0402, 11234014, 11274362 and 11474330) of China and by Hundred-Talent Program and the ``Strategic Priority Research Program(B)" (XDB07000000) of the Chinese Academy of Sciences.

\bibliographystyle{naturemag}
\bibliography{biblio_FeSe_film}

\textbf{Competing Interests} The authors declare that they have no competing financial interests.

\textbf{Correspondence} Correspondence and request for materials should be addressed to Y.-J.S. or H.D.

\textbf{Author contributions} X.-L.P., Z.Q.H. and Y.-J.S. synthesized the samples with assistance from L.L.W and Q.-K.X. P.Z., T.Q. and Y.-L.G. performed the ARPES measurements. P.Z. and X.S. analysed the data. P.R., P.Z., Y.-J.S. and H.D. wrote the manuscript. J.P.H. provided theoretical input. H.D. and Y.-J.S. supervised the project. All authors discussed the paper.

\end{document}